\documentclass[twoside,twocolumn,9pt]{article}
\usepackage{extsizes}
\usepackage[super,sort&compress,comma]{natbib} 
\usepackage[version=3]{mhchem}
\usepackage[left=1.5cm, right=1.5cm, top=1.785cm, bottom=2.0cm]{geometry}
\usepackage{balance}
\usepackage{times,mathptmx}
\usepackage{sectsty}
\usepackage{graphicx} 
\usepackage{lastpage}
\usepackage[format=plain,justification=justified,singlelinecheck=false,font={stretch=1.125,small,sf},labelfont=bf,labelsep=space]{caption}
\usepackage{float}
\usepackage{fancyhdr}
\usepackage{fnpos}
\usepackage[english]{babel}
\addto{\captionsenglish}{%
  
}
\usepackage{array}
\usepackage{droidsans}
\usepackage{charter}
\usepackage[T1]{fontenc}
\usepackage[usenames,dvipsnames]{xcolor}
\usepackage{setspace}
\usepackage[compact]{titlesec}
\usepackage{hyperref}

\usepackage{epstopdf}

\definecolor{cream}{RGB}{222,217,201}

\begin{document}

\pagestyle{fancy}
\thispagestyle{plain}
\fancypagestyle{plain}{

\renewcommand{\headrulewidth}{0pt}
}

\makeFNbottom
\makeatletter
\renewcommand\LARGE{\@setfontsize\LARGE{15pt}{17}}
\renewcommand\Large{\@setfontsize\Large{12pt}{14}}
\renewcommand\large{\@setfontsize\large{10pt}{12}}
\renewcommand\footnotesize{\@setfontsize\footnotesize{7pt}{10}}
\makeatother

\renewcommand{\thefootnote}{\fnsymbol{footnote}}
\renewcommand\footnoterule{\vspace*{1pt}%
\color{cream}\hrule width 3.5in height 0.4pt \color{black}\vspace*{5pt}} 
\setcounter{secnumdepth}{5}

\makeatletter 
\renewcommand\@biblabel[1]{#1}            
\renewcommand\@makefntext[1]%
{\noindent\makebox[0pt][r]{\@thefnmark\,}#1}
\makeatother 
\renewcommand{\figurename}{\small{Fig.}~}
\sectionfont{\sffamily\Large}
\subsectionfont{\normalsize}
\subsubsectionfont{\bf}
\setstretch{1.125} 
\setlength{\skip\footins}{0.8cm}
\setlength{\footnotesep}{0.25cm}
\setlength{\jot}{10pt}
\titlespacing*{\section}{0pt}{4pt}{4pt}
\titlespacing*{\subsection}{0pt}{15pt}{1pt}

\fancyfoot{}
\fancyfoot[LO,RE]{\vspace{-7.1pt}\includegraphics[height=9pt]{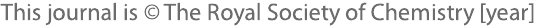}}
\fancyfoot[CO]{\vspace{-7.1pt}\hspace{13.2cm}\includegraphics{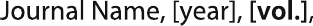}}
\fancyfoot[CE]{\vspace{-7.2pt}\hspace{-14.2cm}\includegraphics{head_foot/RF}}
\fancyfoot[RO]{\footnotesize{\sffamily{1--\pageref{LastPage} ~\textbar  \hspace{2pt}\thepage}}}
\fancyfoot[LE]{\footnotesize{\sffamily{\thepage~\textbar\hspace{3.45cm} 1--\pageref{LastPage}}}}
\fancyhead{}
\renewcommand{\headrulewidth}{0pt} 
\renewcommand{\footrulewidth}{0pt}
\setlength{\arrayrulewidth}{1pt}
\setlength{\columnsep}{6.5mm}
\setlength\bibsep{1pt}

\makeatletter 
\newlength{\figrulesep} 
\setlength{\figrulesep}{0.5\textfloatsep} 

\newcommand{\topfigrule}{\vspace*{-1pt}%
\noindent{\color{cream}\rule[-\figrulesep]{\columnwidth}{1.5pt}} }

\newcommand{\botfigrule}{\vspace*{-2pt}%
\noindent{\color{cream}\rule[\figrulesep]{\columnwidth}{1.5pt}} }

\newcommand{\dblfigrule}{\vspace*{-1pt}%
\noindent{\color{cream}\rule[-\figrulesep]{\textwidth}{1.5pt}} }

\makeatother

\twocolumn[
  \begin{@twocolumnfalse}
\vspace{3cm}
\sffamily
\begin{tabular}{m{4.5cm} p{13.5cm} }

\includegraphics{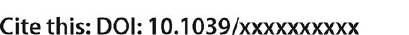} & \noindent\LARGE{\textbf{Lithium ions solvated in helium$^\dag$}} \\
\vspace{0.3cm} & \vspace{0.3cm} \\

 & \noindent\large{Monisha Rastogi,\textit{$^{a}$} Christian Leidlmair,\textit{$^{a}$} Lukas An der Lan,\textit{$^{a}$} Josu Ortiz de Z\'arate,\textit{$^{b}$} Ricardo P\'erez de Tudela,\textit{$^{c}$} Massimiliano Bartolomei,\textit{$^{b}$} Marta I.\ Hern\'andez,\textit{$^{b}$} Jos\'{e} Campos-Mart\'{\i}nez,\textit{$^{b}$} Tom\'{a}s Gonz\'alez-Lezana,$^{\ast}$\textit{$^{b}$} Javier Hern\'andez-Rojas,\textit{$^{d}$} Jos\'{e} Bret\'on,\textit{$^{d}$} Paul~Scheier,\textit{$^{a}$} and Michael Gatchell$^{\ast}$\textit{$^{a,e}$}} \\

\includegraphics{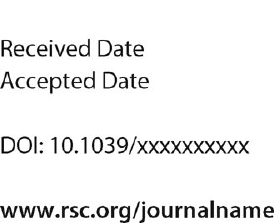} & \noindent\normalsize{We report on a combined experimental and theoretical study of Li$^+$ ions solvated by up to 50 He atoms. The experiments show clear enhanced abundances associated with He$_n$Li$^+$ clusters where $n=2$, 6, 8, and 14. We find that classical methods, e.g.\ Basin-Hopping (BH), give results that qualitatively agree with quantum mechanical methods such as path integral Monte Carlo, diffusion Monte Carlo and quantum free energy, regarding both energies and the solvation structures that are formed. The theory identifies particularly stable structures for $n=4$, 6 and 8 which line up  with some of the most abundant features in the experiments.}

\end{tabular}

 \end{@twocolumnfalse} \vspace{0.6cm}

  ]

\renewcommand*\rmdefault{bch}\normalfont\upshape
\rmfamily
\section*{}
\vspace{-1cm}


\footnotetext{\textit{$^{a}$~Institut f\"{u}r Ionenphysik und Angewandte Physik, Universit\"{a}t Innsbruck, Technikerstr.~25, A-6020 Innsbruck, Austria. ; E-mail: michael.gatchell@uibk.ac.at}}
\footnotetext{\textit{$^{b}$~Instituto de F\'{\i}sica Fundamental, IFF-CSIC, Serrano 123, 28006 Madrid, Spain.}}
\footnotetext{\textit{$^{c}$~Lehrstuhl f\"ur Theoretische Chemie, Ruhr-Universit\"at Bochum, 44780 Bochum, Germany.}}
\footnotetext{\textit{$^{d}$~Departamento de F\'{\i}sica and IUdEA, Universidad de La Laguna, 38205 Tenerife, Spain.}}
\footnotetext{\textit{$^{e}$~Department of Physics, Stockholm University, 106 91 Stockholm, Sweden.}}

\footnotetext{\dag~Electronic Supplementary Information (ESI) available: Radial distributions of He$_n$Li$^+$ snowballs.. See DOI: 10.1039/b000000x/}



\section{Introduction}
Ultra-cold helium nanodroplets have on numerous occasions proven to be a powerful tool for studying systems ranging from individual atoms to complex nanoparticles.\cite{Stienkemeier:2006aa,Mauracher:2018aa} Despite their low temperature (0.37\,K) and weak interactions, they are remarkably good at solvating dopants, which typically reside near the core of the droplets and can, for example, be utilized for assembling clusters of one or more species. An interesting exception to this are the alkali metals that, at least in the case of individual atoms and small clusters, are ``heliophobic'' and reside in dimples on the surface of the droplets.\cite{Ancilotto:1995aa,Stienkemeier:1996aa,Nakayama:2001aa} This is due to a balance between the Pauli repulsion between the unpaired valence electron in the metal atoms and the closed shell He electrons on the one hand, and the surface tension of the droplet on the other. In contrast to \emph{neutral} alkali metal atoms, alkali \emph{cations} interact very strongly with He and are readily solvated, forming several layers of solvation shells around the ions. For the innermost layers, the He density can surpass the density of solid He, giving structures known as Atkins snowballs.\cite{Atkins:1959aa}

Several theoretical and experimental studies have focused on the solvation structures of He snowballs around alkali metal cations. Reatto \emph{et al.}\ employed variational Monte Carlo (MC) simulations with shadow wave functions to probe alkali ion impurities (Li$^+$, Na$^+$, K$^+$, and Cs$^+$) in liquid helium for equilibrium densities at 0\,K.\cite{Buzzacchi:2001aa,Galli:2001aa,Rossi:2004aa} The chemical potential, local order, single particle excitation, and effective ionic mass were determined from the simulation model. A substantial difference in the snowball for corresponding ions could be observed. It was predicted that only Na$^+$ and K$^+$ have a tendency to form a solid snowball whereas the localization is not as prominent for Li$^+$ and Cs$^+$ species. Gianturco \emph{et al.}\ later conducted a dedicated study based on the solvation of Li and other alkali metals in helium matrices employing a combination of classical energy minimization techniques and of exact quantum Diffusion Monte Carlo (DMC) methods.\cite{Di-Paola:2005aa,CBMGYYY:JCP07,Sebastianelli:2006aa,MCBGYYY:TCA07} Small He$_n$Li$^+$ clusters with $n \leq 30$ were considered for their investigations and they treated the full cluster interaction as a sum of pairwise potentials for Li$^+$--He and He--He. It could be deduced that three particularly stable structures exist at $n = 6$, 8, and 10 with the most stable structure being found for $n = 6$. Additionally, evaluation of single particle evaporation energies, employing classical and quantum techniques, shed light on the formation of a rigid layer of helium with approximately 8 atoms being more tightly bound to the central ion. After this first shell, the evaporation energy was mainly governed by He--He interaction and not by interactions with the ionic core. The behavior was found to be similar to Na$^+$ and K$^+$ doped helium clusters, where the initial rigid layer was comprised of 9 and 12 He atoms, respectively. This rigid behavior of a fully developed first solvation shell for $n=8$ was also reported in the ground state path integral calculation performed by Paolini \emph{et al.}\ \cite{PAT:JCP07} who found a stable structure of He atoms forming two parallel squares rotated by $\pi/4$ with respect to each other repeated in successively larger clusters ($n \sim 70$, for example). 

Previous investigations found that three-body (3B) contributions are rather insignificant in the stability of these helium clusters doped with alkali ions.
Marinetti \emph{et al.}\cite{MCBGYYY:TCA07} observed some shifts of the radial distributions to slightly larger distances
in their study on He$_n$Li$^+$ when the coupling between induced dipoles on the He atoms were taken into account. 
The \emph{ab initio} calculations of the potential energy curve of HeLi$^+$ and optimal structures for He$_n$Li$^+$ with $n=1$--6 performed by
Sebastianelli \emph{et al.}\cite{Sebastianelli:2006aa} concluded that the overall interactions were in fact governed mainly by diatom-like interactions between the ion and He atoms and that the pairwise approximation turns to be an acceptable description for these systems.
The theoretical investigations performed by Issaoui \emph{et al.}\cite{IAGYCO:JCP14} to study He$_n$Na$^+$ clusters added a self consistent many-body contribution between induced dipoles to the pairwise diatomic energy curves.
The analysis performed on these studies revealed a notable overestimation of the energies predicted by the 2B approximation in larger clusters, which was found to delay the onset of delocalization and snowball features.\cite{IAGYCO:JCP14}

M{\"u}ller \emph{et al.}\ carried out a systematic investigation on the formation and stability of helium snowballs created by employing femtosecond photoionization (PI) and electron impact ionization (EII) of alkali clusters (Na, K, Rb and Cs).\cite{Muller:2009ab} From PI spectra, it could be deduced that alkali metal ions that originate from fragmented alkali clusters are more likely to constitute snowball complexes than their ionized monomer counterparts. This could be attributed to the fragmentation of clusters into singly charged ions, due to multiple ionization. Additionally, it was concluded that the size of a snowball with respect to the mass of alkali metals is a function of the kinematics of photofragmentation. For Na$^+$ and K$^+$ ions, they only observed the formation of small snowball sizes (up to 3 and 10 He atoms, respectively), which prohibited a direct comparison with predicted first shell closures from theory. However, with the heavier Rb$^+$ and Cs$^+$ ions they observed the formation of snowballs with up to 41 He atoms and identified the closures of the first solvation layers at He$_{14}$Rb$^+$ and He$_{16}$Cs$^+$, somewhat smaller than the shell sizes predicted by theory.\cite{Muller:2009ab,Rossi:2004aa} Later, An der Lan \emph{et al.}\ studied He droplets containing Na and K monomer and dimer cations.\cite{Lan-Lukas:2012aa} They reported on snowballs containing up 30 He atoms, with the first shell closures identified after He$_9$Na$^+$ and He$_{12}$K$^+$. The lightest alkali ion, Li$^+$, was excluded from both of these experimental studies (and others like them) as the small mass and isotopic composition of the Li ion could potentially obstruct the evaluation of mass spectrometric data, corresponding to alkali-helium snowball complexes.

In this present work we report on the solvation of Li$^+$ ions in helium, evaluated with high-resolution mass spectrometry measurements and different theoretical methods. In our experiments, He$_n$Li$^+$ complexes containing several tens of He atoms are identified and anomalies in specific cluster size yields let us probe the ion stabilities of these systems. These results are compared with both classical and quantum mechanical (QM) simulations of a Li$^+$ ion solvated with He atoms. In particular, and in a similar fashion as previous investigations of clusters formed doping coronene molecules with rare gas atoms and H$_2$ \cite{RPBHCGVHB:JCP15,RBHCGVPPHB:JCP17,BPGHCVHBP:JCP17}, we have carried out basin-hopping (BH), DMC and path integral Monte Carlo (PIMC) calculations. In addition to this, estimations of the quantum free energy (QFE) have been calculated, leading to very similar results to those obtained with QM corrections of the BH results including zero-point energy (ZPE) effects. Geometries and energies of the stable configurations observed for the different He$_n$Li$^+$ clusters have been investigated and, in particular, the behavior as a function of the size of each cluster has been analyzed in an attempt to understand the abundances observed in the experiment for each $n$.

The structure of the paper is as follows: In Section \ref{sec_exp} we present the essential details of the experimental setup, in Section \ref{theo} we present the theoretical approaches employed in this work, and in Section \ref{sec_resul} results are shown and discussed. Finally in Section \ref{sec_conclu} the conclusions are listed.

\section{Experimental details}
\label{sec_exp}

Pure, pre-cooled helium (purity 99.9999\%) with a stagnation pressure of 25\,bar, was expanded through a 5\,$\mu$m nozzle cooled to 6.5\,K, leading to the formation of droplets with a broad size distribution (averaging about 10$^7$--10$^8$ He atoms). The resulting supersonic beam passed through a conical skimmer (diameter 0.8\,mm) which is located 8\,mm downstream from the nozzle. The skimmed beam travels across a 20\,cm long differentially pumped pick up region where it is doped with high purity lithium (99\% trace metals basis from Sigma Aldrich). The lithium sample was introduced into a cylindrical pickup cell under an inert atmosphere and covered with hexane to prevent oxidation during the transfer to the vacuum chamber. After vaporization of the hexane by evacuation at room temperature, the pickup-cell was resistively heated to a temperature of 750\,K. The doped helium droplets were ionized by electron impact with kinetic energies of 70\,eV. Resulting cations were then extracted from the ion source and guided into the extraction region of a commercial reflectron time-of-flight mass spectrometer (Tofwerk AG, model HTOF, mass resolution $\Delta m/m=1:5000$). Additional experimental details have been described elsewhere.\cite{An-der-Lan:2011aa,Leidlmair:2011aa}

\section{Theoretical methods}
\label{theo}

\subsection{Potential energy surface}
\label{sec_pes}

The employed force field is based on the sum of two-body (2B) He-Li$^+$ and
He-He non-covalent interaction contributions.
 For the He-He interaction we use the potential reported in Ref.\cite{LDLCCASTP:CS08}
while for the He-Li$^+$ contribution we have developed a new
 potential based on accurate CCSD(T) results obtained in the complete
basis set limit. In both cases the adopted analytical representation
exploits the improved Lennard Jones (ILJ) formulation given by \cite{PBRCCV:PCCP08}:

\begin{equation}
\label{ILJ}
V(r) = \epsilon 
\left[
{\frac{m}{n(r) \! - \!m}} \left( \frac{r_m}{r} \right)^{n(r)}
\! - {\frac{n(r)}{n(r) - m}} \left(\frac{r_m}{r} \right)^{m}
\right]
\end{equation}
\noindent 
where $\epsilon$ is the potential depth, $r_m$ the position of the minimum and $n(r)$ is defined as follows \cite{PBRCCV:PCCP08}: 

\begin{equation}
\label{nILJ}
n(r) = \beta + 4 \left( \frac{r}{r_m} \right)^2.
\end{equation}

The corresponding parameters for both the He--He and He--Li$^+$ potentials using the ILJ analytical expression are given in Table \ref{table1}. The effects of 3B terms are investigated by introducing an induced dipole-induced dipole interaction as that employed in previous studies  \cite{LWHS:JCP16,MCBGYYY:TCA07} with damping functions in our PES:

\begin{eqnarray}
\label{eq:3}
V_{\rm 3B} & = & -\alpha^2
\left[
{\frac{3 r_j}{4}} g_3(r_i) g_5(r_{ij}) + {\frac{3 r_i}{4}} g_3(r_j) g_5(r_{ij}) \right. \nonumber \\ 
& - & {\frac{1}{4}} g_3(r_i) g_3(r_j) g_1(r_{ij}) 
-{\frac{3}{2}} g_1(r_i) g_1(r_j) g_5(r_{ij}) \nonumber \\ 
& - & \left.
{\frac{1}{2}} g_1(r_i) g_3(r_j) g_3(r_{ij}) - {\frac{1}{2}} g_3(r_i) g_1(r_j) g_3(r_{ij})
\right]
\end{eqnarray}
\noindent
where $\alpha=1.31\, a_0^3$ for the He polarizability, $r_i$ and r$_j$ are He--Li$^+$ distances, $r_{ij}$ is the He--He distance, and $g_n(r_i) = f_n(r_i)/r_i^n$, where $f_n(r)$ are the damping functions expressed as \cite{SLLMW:CPL01}:

\begin{equation}
f_n(r) = 1 - \exp(-br) \sum_{k=0}^n {\frac{\left[br \right]^k}{k!}},
\end{equation}
\noindent 
with $b$ being equal to 2.9\,$a_0^{-1}$ or 3.2\,$a_0^{-1}$ for the He--Li$^+$ and He--He interaction, respectively. The 3B calculation has been performed with a value of $\beta = 9$ in the ILJ description of the He--He interaction.

\begin{table}[h]
\begin{center}
\small
  \caption{\ Parameters for the ILJ potentials for the He--Li$^+$ and He--He interactions. $r_m$ is given in \AA, $\epsilon$ in meV; $m$ and $\beta$ are dimensionless.}
\label{table1}
\begin{tabular*}{0.5\textwidth}{@{\extracolsep{\fill}}lcccc}
  \hline
     & $m$ & $r_m$ & $\epsilon$ 
      & $\beta$ 
      \\
    \hline
   He--Li$^+$ & 4 &  1.90  &  81.3 & 4.2    \\ 

 He--He & 6 & 2.97 &  0.947  & 8      \\   
    \hline
  \end{tabular*}
\end{center}
\end{table}

\subsection{Basin-Hopping}
\label{theo_BH}
The BH \cite{WD:JPCA97} is a stochastic method to obtain the global minima of a potential energy surface (PES). This technique transforms the surface into a collection of basins which are explored by hopping between the local minima. 
Both local and global minima are preserved under this transformation.
A Metropolis criterion using the energies of the initial and final minima in each step at a fictitious temperature determines if the attempted steps are accepted or rejected. The algorithm is particularly efficient since downhill barriers between different basins are removed and trapping is usually avoided.
Moreover, size steps are typically larger than those employed for thermal sampling in MC simulations. 
The calculation was performed with a constant fictitious temperature such that $k_B T = 1.5$\,meV.
The BH approach has been successfully employed for a large series of molecular clusters \cite{WD:JPCA97,HCBG:JPCC12,ABLGH:JCP12,HBGW:JPCB06,DW:PRB99,RW:JCP03,HBGW:JCP04,HBGW:CPL05,HCRBG:JPCA10,W:Book03}, and in particular, it has been an extremely useful tool in recent investigations of coronene doped with rare gas atoms and molecular hydrogen.\cite{RPBHCGVHB:JCP15,RBHCGVPPHB:JCP17,BPGHCVHBP:JCP17}
Quantum effects can be included by means of the ZPE in the harmonic approximation 
\cite{RPBHCGVHB:JCP15,BPGHCVHBP:JCP17}, a calculation which requires to construct a database of local minima close to the global minimum for each cluster size.

\subsection{Quantum Free Energy}
\label{theo_freeE}
The QFE of a specific minimum $\alpha$ at a temperature $T$ of a cluster with $n$ He atoms is given by:

\begin{equation}
\label{freeE}
F_\alpha (T) = -k_B T \ln  Z_\alpha (T)
\end{equation}
\noindent
where $k_B$ is the Boltzmann constant and $Z_\alpha (T)$ is the partition function of the minimum $\alpha$ at constant temperature $T$.
Under a harmonic approximation, this partition function can be written as \cite{CDW:JCP01}:

\begin{equation}
\label{Part}
Z_\alpha (T) = {\frac{2 n!}{{\cal O}_\alpha}} e^{-\beta E_\alpha}
\prod_i {\frac{e^{-\beta \hbar \omega_i^\alpha / 2} }{ 1 - e^{-\beta \hbar \omega_i^\alpha}}}
\end{equation}
\noindent
where ${\cal O}_\alpha$ is the order of the point group corresponding to the minimum $\alpha$, $\hbar$ is the Planck constant, $\beta = (k_B T)^{-1}$, and 
$\omega_i^\alpha$ is the $i$-th vibrational frequency associated with the minimum with energy $E_{\alpha}$.
In the present calculations we consider $T=2$\,K.

There is an alternative version of this method in which the partition function given in Eq. (\ref{Part}) is replaced by its classical expression. This classical free energy tends asymptotically to the BH results when the temperature is decreased to zero, whereas the QFE tends to the QM corrected BH+ZPE values.    

\subsection{Diffusion Monte Carlo}
\label{theo_DMC}

QM calculations were carried out by means of the DMC method.
In this algorithm, the time-dependent Schr{\"o}dinger equation is transformed into a diffusion
equation after substitution of the real time $t$ by the imaginary time $\tau= i t$. The ground
state wave function can be then obtained from the longest lasting term ($\tau \rightarrow \infty$)
in the solution of the diffusion equation. Details of the method can be found elsewhere.\cite{A:JCP75,SW:PR91,RPBHCGVHB:JCP15}
Ground state energies and probability densities were computed using a code developed by Sandler
and Buch\cite{B:JCP92,SB:PC99}, assisted with the descendant weighting method. For a given cluster size,
six simulations were typically performed, each of them involving nine generations for the descandant
weighting procedure. About 12000 replicas were propagated with time steps ranging from 40 to 80 a.u.\
and for around 6000 steps. The initial set of replicas consisted in Gaussian spatial distributions
(widths between 0.3-0.4 \AA) around the classical equilibrium cluster geometry. It was found that
the calculations can run optimally if the initial distribution is obtained by scaling the
equilibrium geometries by a factor of 1.1--1.5.

\begin{figure*}[]
   \centering
   \includegraphics[width=1\textwidth]{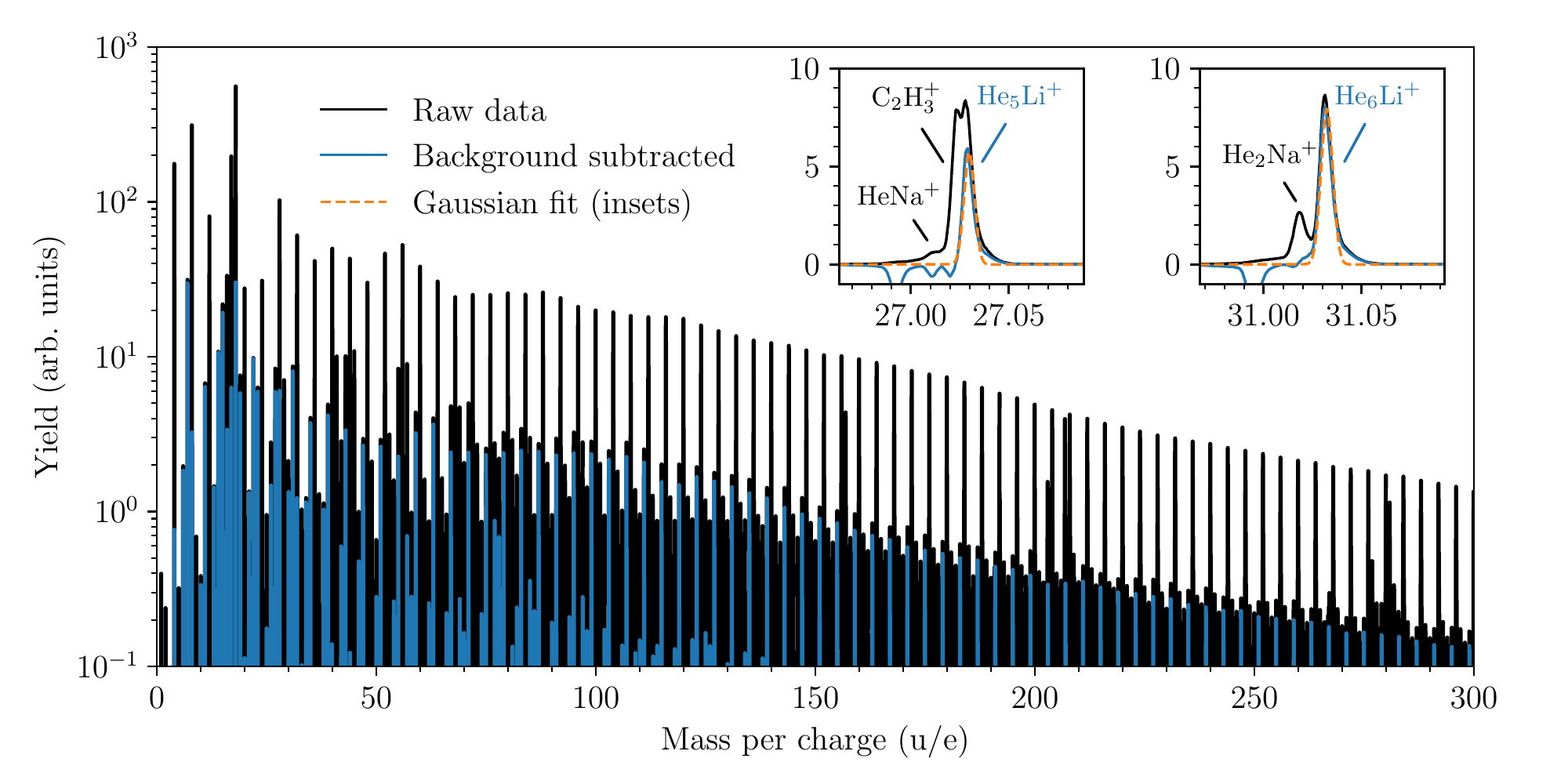} 
   \caption{Mass spectrum from helium nanodroplets doped with lithium. The black spectrum is the raw spectrum as measured and the blue spectrum is what remains following background subtraction (mainly ions containing Li). The background spectrum was measured with identical experimental conditions except that the Li-containing oven was operated at a lower temperature. This lower temperature is sufficient to vaporize the main pollutants in the sample (e.g.\ Na), but not Li. Gaussian fits (orange dashed lines in insets) are performed for each He$_n$Li$^+$ peak.} 
   \label{fig:MS}
\end{figure*}

\subsection{Path Integral Monte Carlo}
\label{theo_PIMC}

The PIMC method has been described in detail before \cite{PMGRMDV:JCP10,RGV:IRPC16} and therefore here we will restrict ourselves to give the most relevant aspects for the present calculation. 
In essence this approach is based on the expression of the density matrix at a temperature $T$ 
as the product of $M$ density matrices at a higher-temperature $T' = T \times M$.
The density matrix is therefore evaluated in a collection of quantities
${\cal R}_{\alpha} \equiv \left\{ {\bf r}_1^{\alpha},\ldots,{\bf r}_N^{\alpha}\right\}$, where $\alpha$ runs over the $M$ quantum beads, containing the position vectors ${\bf r}_i^{\alpha}$ of the $N$ particles which form the cluster: the Li$^+$ and the $n$ He atoms. 
In particular, the energy of each cluster can be obtained by means of the \emph{thermodynamic} estimator developed by Baker \cite{B:JCP79}:

\begin{equation}
\label{Ethermo}
\langle E \rangle_{\rm thermo} 
= {\frac{3 {N} }{ 2 \tau}} - \left \langle 
\sum_{\alpha =0}^{M-1} \sum_{i=1}^{N}
{\frac{({\bf r}_i^{\alpha}-{\bf r}_i^{\alpha+1})^2}{ 4 M \lambda_m \tau^2}}
- V,
\right \rangle
\end{equation}
\noindent
with $\lambda_m = \hbar^2 / 2m$, $m$ being the mass of either He or Li$^+$ and $\tau = \beta / M$.
The first term in Eq. (\ref{Ethermo}) corresponds to the classical kinetic energy multiplied by $M$ and the average of the energy due to the
spring-like interaction assumed between consecutive beads in the 
same ring describing a specific particle and
the potential energy $V$ is performed over the MC steps.
The PIMC calculation has been performed at 2\,K using $M = 200$ beads, which are moved in groups of 10 following a staging method.\cite{C:RMP95,PC:PRB84}

\section{Results and discussion}
\label{sec_resul}

\subsection{Experimental Results}
\label{disc_exp}
A cationic mass spectrum from helium nanodroplets doped with lithium is shown in Fig.\ \ref{fig:MS}. The raw mass spectrum (in black) is dominated by the pure He$_n^+$ cluster series and by subtracting a background measurement we can largely isolate the products containing lithium (in blue). The background measurement is performed under identical experimental conditions as the main experiment, but with the lithium-containing oven operating at a lower temperature ($< 600$\,K). This lower temperature is insufficient for vaporizing Li, but is sufficient for vaporizing Na, which is the main pollutant in our lithium sample and forms He snowballs of similar masses. From the reduced spectrum (blue in Fig.\ \ref{fig:MS}) we deduce the abundances of He$_n$Li$^+$ snowballs by fitting each peak with a Gaussian profile, two examples with particularly large overlaps with other peaks are shown in the insets. Beyond the He$_n$Li$^+$ series we observe pure Li$_n^+$ clusters for $n = 2$ and 3, but no larger clusters irrespective of the pickup conditions (even with an oven temperature as high as 1100\,K).

\begin{figure}[t]
   \centering
   \includegraphics[width=0.95\columnwidth]{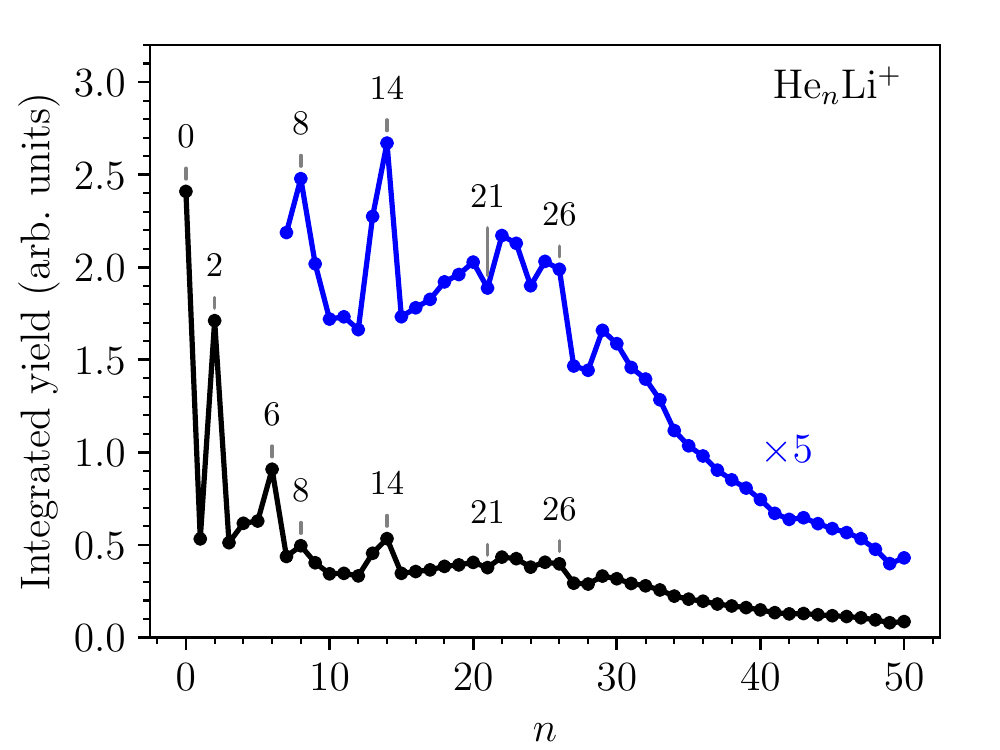} 
   \caption{Measured abundances of He$_n$Li$^+$ as a function of number of He atoms, $n$. Maxima are observed for $n=2$, 6, 8, and 14, while distinct minima over the underlying size distribution are seen for $n=21$, 24, 27, and 28. The statistical errors of the integrated yields are smaller than the data markers.}
   \label{fig:Ints}
\end{figure}

The integrated counts from each He$_n$Li$^+$ complex are shown in Fig.\ \ref{fig:Ints}. The standout features in this spectrum are the local maxima observed for $n=2$, 6, 8, and 14, indicating that these are particularly stable systems (compared to their neighbors). At higher masses there are a few dips in the spectrum at $n=21$, 24, and 27--28 on top of an underlying distribution that smoothly tapers off towards large cluster sizes.

\subsection{Theoretical Results}
\label{disc_theo}

\begin{figure}[t]
   \centering
   \includegraphics[width=0.95\columnwidth]{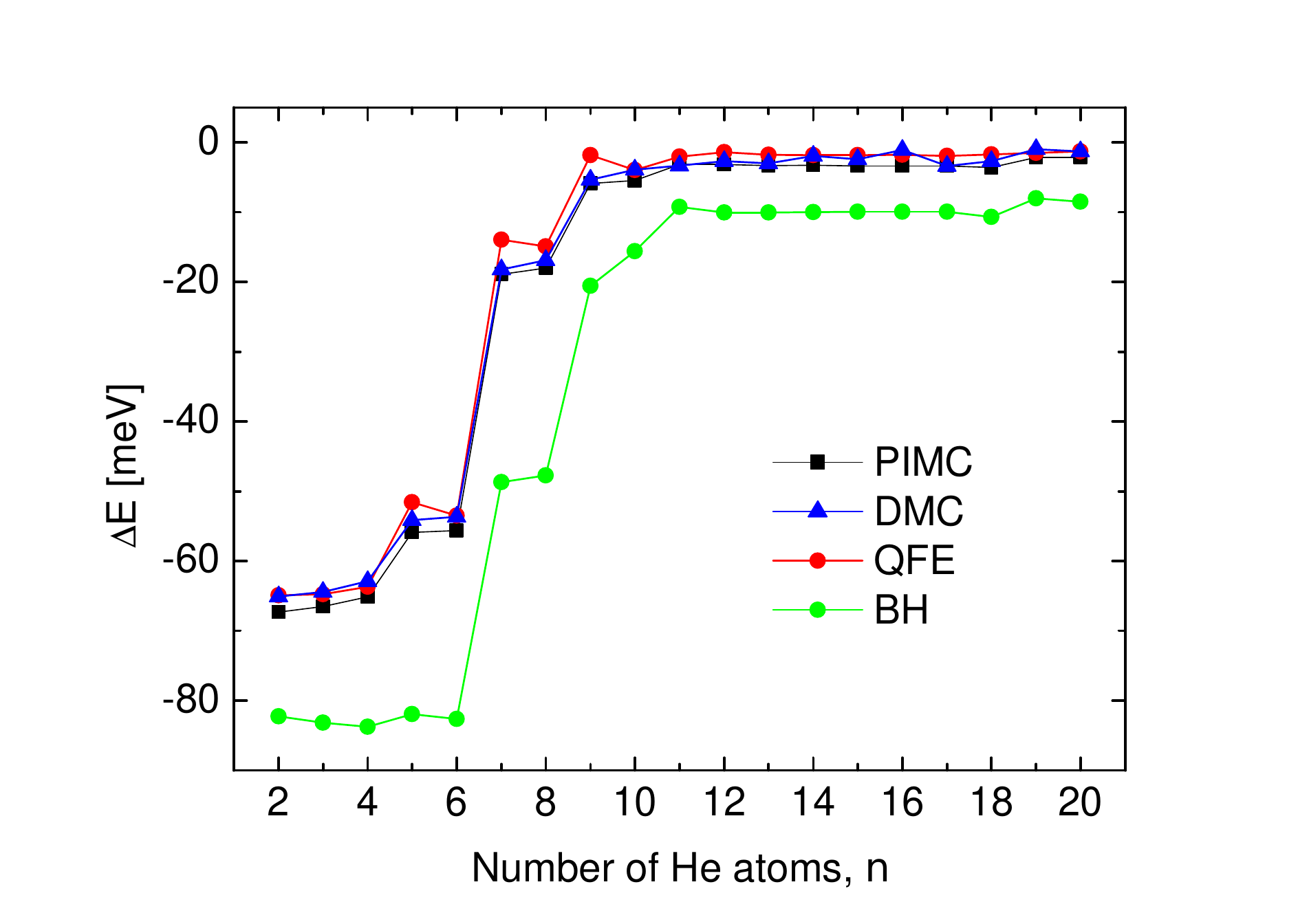} 
   \caption{Evaporation energies obtained by means of the BH (green circles), QFE (full red circles), DMC (blue triangles) and PIMC (black squares) methods.} 
   \label{fig:Eevap}
\end{figure}

From a theoretical point of view, we have investigated the behavior of some quantities as a function of the number of He atoms $n$ in the cluster, searching for features which may point out the stability of specific sizes of He$_n$Li$^+$ consistent with the measured abundances shown in Fig. \ref{fig:Ints}.
Thus, in Figure \ref{fig:Eevap} we show the evaporation energies, $\Delta E = E_n-E_{n-1}$, obtained by means of those methods discussed in Section \ref{theo}.
Despite the expected quantitative differences, the classical approach we show here, the BH method (see Section \ref{theo_BH}) exhibits qualitatively the same trend as the corresponding QM counterparts employed, that is, QFE, DMC and PIMC (discussed in \ref{theo_freeE}, \ref{theo_DMC} and \ref{theo_PIMC}, respectively).
Results obtained with the QM corrected version of the BH approach, the BH+ZPE method, are almost identical to the QFE energies and are not included in the figure. 
The main abrupt changes observed in the corresponding energy curves as a function of the size of the cluster occur at $n = 4, 6$ and 8.
Jumps in the evaporation energies for these same sizes were also reported in previous investigations of the system \cite{Di-Paola:2005aa,CBMGYYY:JCP07,PAT:JCP07} and were interpreted as indications of the presence of stable structures for He$_6$Li$^+$ and He$_8$Li$^+$.\cite{CBMGYYY:JCP07}

\begin{figure}[t] 
   \centering
   \includegraphics[width=0.95\columnwidth]{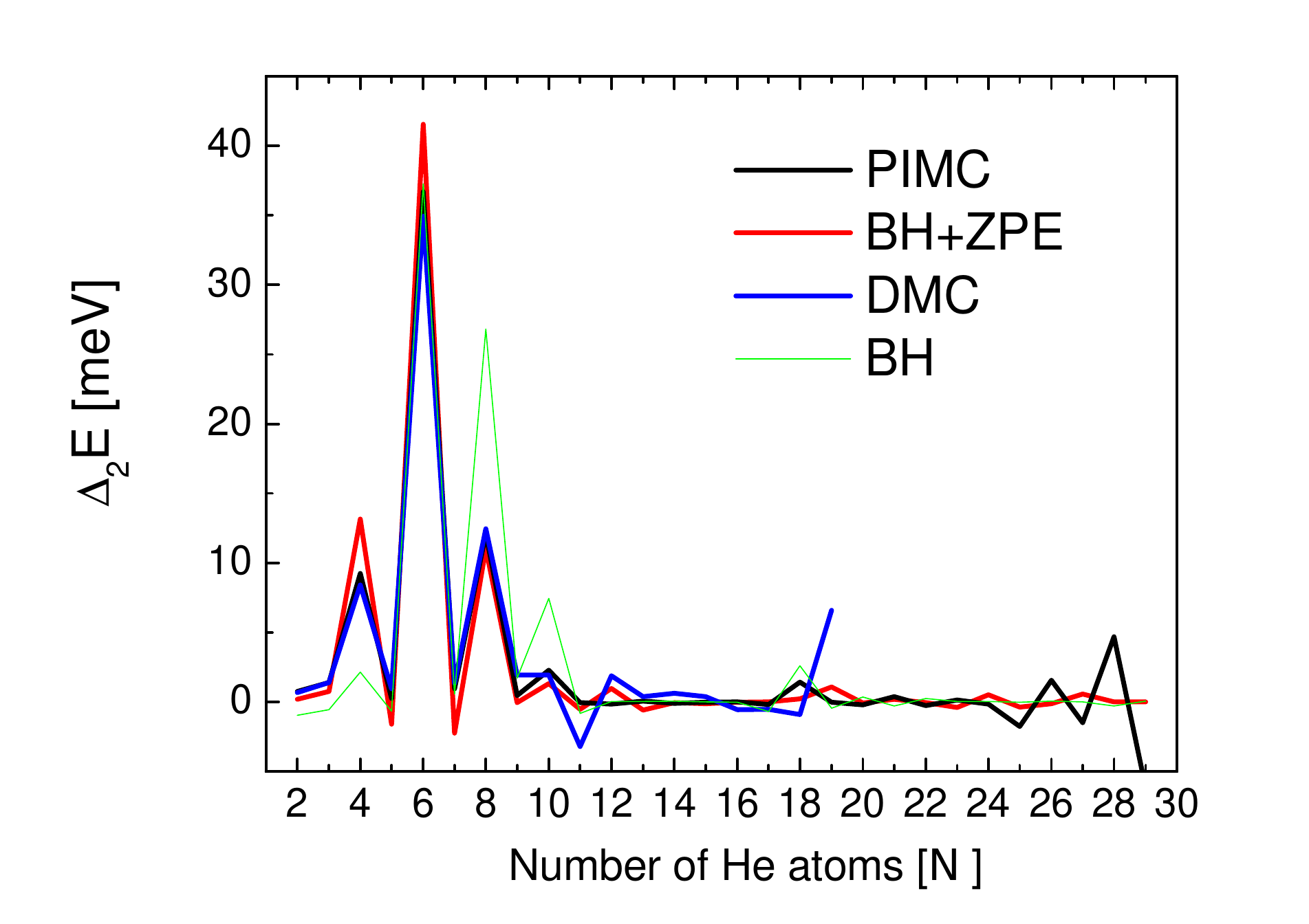} 
   \caption{Second difference energies (see text for details) obtained with the PIMC (black), QFE (red), BH+ZPE (blue) and BH (green) theoretical methods.} 
   \label{fig:2ndE}
\end{figure}

\begin{figure*}[t]
   \centering
   \includegraphics[width=0.70\textwidth]{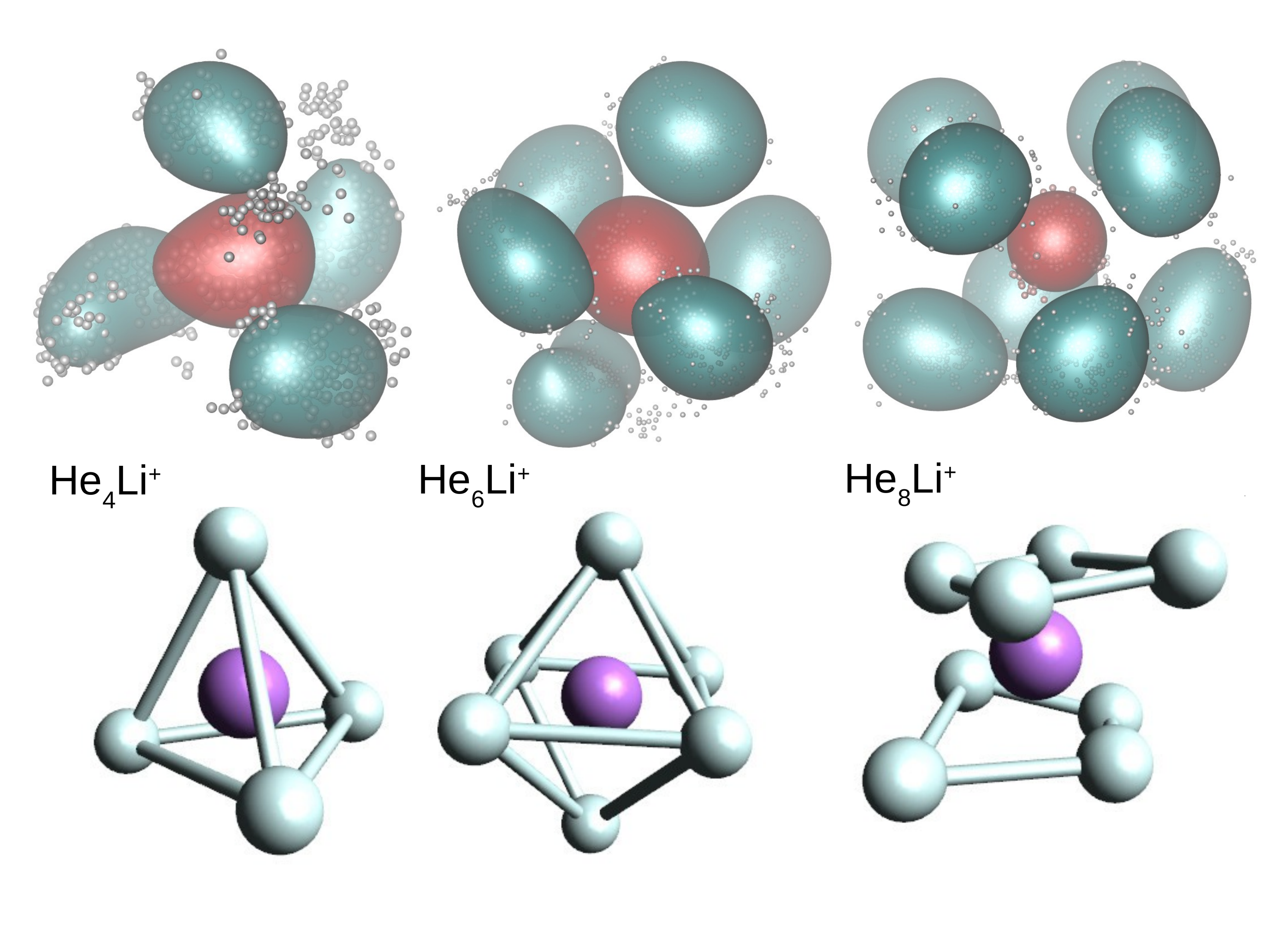} 
   \caption{Probability density functions for He$_4$Li$^+$ (left), He$_6$Li$^+$ (middle) and He$_8$Li$^+$ (right) obtained by means of the PIMC (top) and DMC (bottom) methods. Green color in the PIMC results are for the average location of the He atoms and red for the Li$^+$ atom. The average position of He (Li$^+$) atoms in the replica of the DMC distributions are in white (purple). See text for details.} 
   \label{fig:dist}
\end{figure*}

Second energy differences, defined as $\Delta_2 E =  E_{n+1} + E_{n-1} - 2 E_n$, are also a useful magnitude to search for stable He$_n$Li$^+$ clusters at specific numbers of He atoms. The results obtained by means of the PIMC, BH+ZPE and QFE approaches are shown in Figure \ref{fig:2ndE}. 
As expected the features observed in the curve of the evaporation energy of Figure \ref{fig:Eevap} also manifest as peaks when we plot these $\Delta_2 E$ differences. Thus, noticeable maxima are observed also at $n=4$, 6, and 8, which, in view of the BH result also included in Figure \ref{fig:2ndE}, seem to have their origin in the minima of the PES. The comparison between the second energy differences obtained by means of the BH and those calculated with the other methods reveals however noticeable discrepancies between classical and QM approaches.
The classical result suggests a similar feature at $n=10$ as well but the QM calculations do not entirely confirm this regard, in apparent agreement with the experiment. 
The integrated yield shown in Figure \ref{fig:Ints} also exhibits maxima at $n=6$ and 8, whereas for $n=4$, only a suggested shoulder is seen.

Two of the most prominent maximum peaks observed in Figure \ref{fig:2ndE}, those for He$_6$Li$^+$ and He$_8$Li$^+$, are also clearly visible in the measured abundances shown in Figure \ref{fig:Ints}. 
The equilibrium geometries associated to the minimum energy configurations have been investigated before in previous works: He$_6$Li$^+$ exhibits a symmetrical octahedral configuration with the He atoms coordinating the Li$^+$ impurity, located at the center of the cluster.\cite{Di-Paola:2005aa,MCBGYYY:TCA07}
He$_8$Li$^+$, on the other hand, has a stable configuration formed by two parallel squares rotated by $\pi/4$ to each other surrounding the Li$^+$ ion \cite{PAT:JCP07}, which was
found also as the inner core of larger clusters, thus suggesting that it corresponds to the 
geometry of the first solvation shell. 
This hypothesis was confirmed by the integration of the shell performed by Paolini \emph{et al.}\ \cite{PAT:JCP07} with ground state path integral calculations which yielded a value of 8.24 He atoms.

In this work we have performed DMC and PIMC calculations of the probability density functions corresponding to the He$_n$Li$^+$ clusters with $n=4$, 6 and 8, those which correspond to special features in the curves as a function of the number of He atoms shown in Figures \ref{fig:Eevap} and \ref{fig:2ndE}. 
In the top panels of Figure \ref{fig:dist} we show the PIMC distributions obtained using a representation on the Eckart frames for specific snapshots of the quantum beads for each atom and their corresponding average represented as a cloud surrounding the expected location of both the He and Li$^+$ atoms. The choice of a system satisfying Eckart conditions\cite{Eckart:1935aa} to guarantee an optimal separation between rotation and vibration, is made here only for pictorial purposes.
Analogously, geometries obtained by averaging the positions of the DMC replicas (after rotation to a common body-fixed frame) are included in the figure.
Both methods yield distributions which are not far from the equilibrium structures predicted by classical energy minimization algorithms.\cite{Di-Paola:2005aa}
Thus, the QM approaches find a structure for He$_4$Li$^+$ which contains the ionic impurity caged inside a tetrahedron formed by the four He atoms and, for $n=6$ and 8, the above mentioned octahedral and parallel squares structures found in previous investigations are reproduced here
by means of the DMC and PIMC calculations. 
Although the probability density functions (not shown here) for the inter-particle distances, He--He and He--Li$^+$, and the corresponding angles obtained with the DMC approach, certainly exhibit an intrinsic broadening, the maxima are only slightly deviated with respect to the stick values predicted in classical energy minimization studies.

\begin{figure}[t]
   \centering
   \includegraphics[width=0.95\columnwidth]{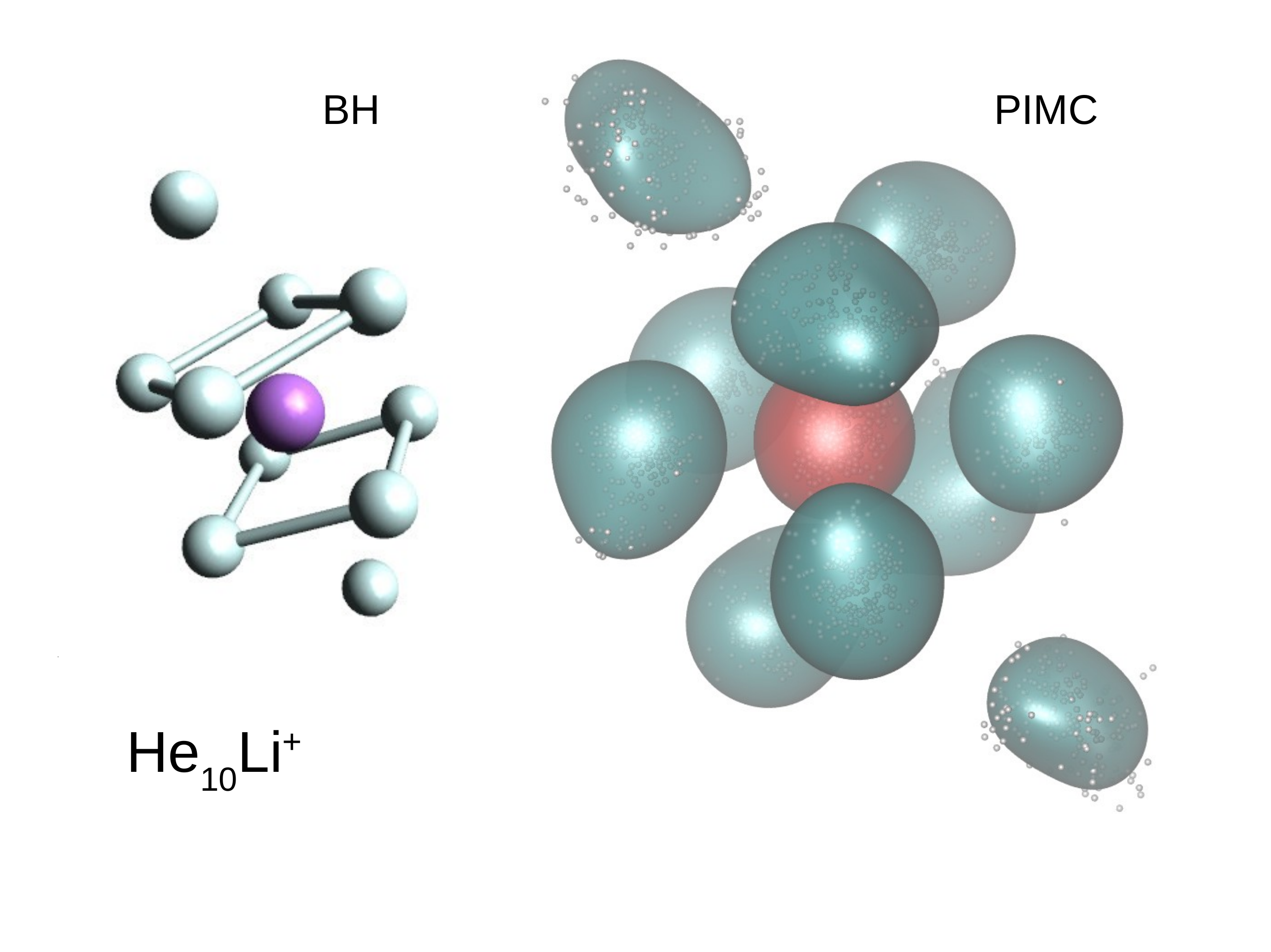} 
   \caption{BH optimized geometry  (left) and PIMC probability density function (right) for He$_{10}$Li$^+$.} 
   \label{fig:n10}
\end{figure}

Our calculations also reveal the stability of the structure found for He$_{8}$Li$^+$. Thus, 
Figure \ref{fig:n10} shows BH and PIMC results for He$_{10}$Li$^+$ indicating that both the classical optimized geometry and the QM probability density function consists, in essence, of the core observed at $n=8$ with the two extra He atoms located over the center of each parallel square. This result is consistent with previous findings for this particular cluster.\cite{Di-Paola:2005aa}
This trend is maintained even for larger cluster sizes, and the analysis of radial and angular distributions reveals that the inner shell is quite similar to the structure seen for He$_{8}$Li$^+$ (see supplementary information). 

The comparison between these theoretical results and the experimental abundances reveals agreement for peaks at $n=6$ and 8. However, the prominent maximum seen for $n=2$ in the experimental data in Figure \ref{fig:Ints} does not have a definitive direct explanation from the theory. This abundance anomaly from $n=2$ is also observed in experiments with Na$^+$ and K$^+$ ions\cite{Lan-Lukas:2012aa} which suggests that this is a product of the ionization mechanism itself, which is not covered by the simulations. One possible explanation is that a He$_2^*$ is formed by the initial electron impact which then through associative Penning ionization forms a He$_2$Li$^+$ complex. Furthermore, the high abundance of the He$_{14}$Li$^+$ complex in the experiments is not reproduced by calculations. This structure could be explained by the nesting of a parallel square structure like He$_8$Li$^+$ in an octahedron like He$_6$Li$^+$, or vice versa, similar to the the nested solvation shells observed for the He$_n$Ar$^+$\cite{Bartl:2014aa} and H$_n^-$\cite{Renzler:2016aa} clusters. However, a particularly stable structure with such a geometry is not observed in the present simulations. 
In fact, further DMC calculations were carried out starting with a geometry where a He$_6$Li$^+$ octahedron is nested inside a cube formed with eight He atoms (a higher energy classical local minimum) but, after the simulations, the cluster rearranged to a structure with a core formed by eight atoms. 
This final geometry was not particularly stable as compared with their closest neighbors $n=13$ and 15.

\begin{figure}[t] 
   \centering
   \includegraphics[width=0.95\columnwidth]{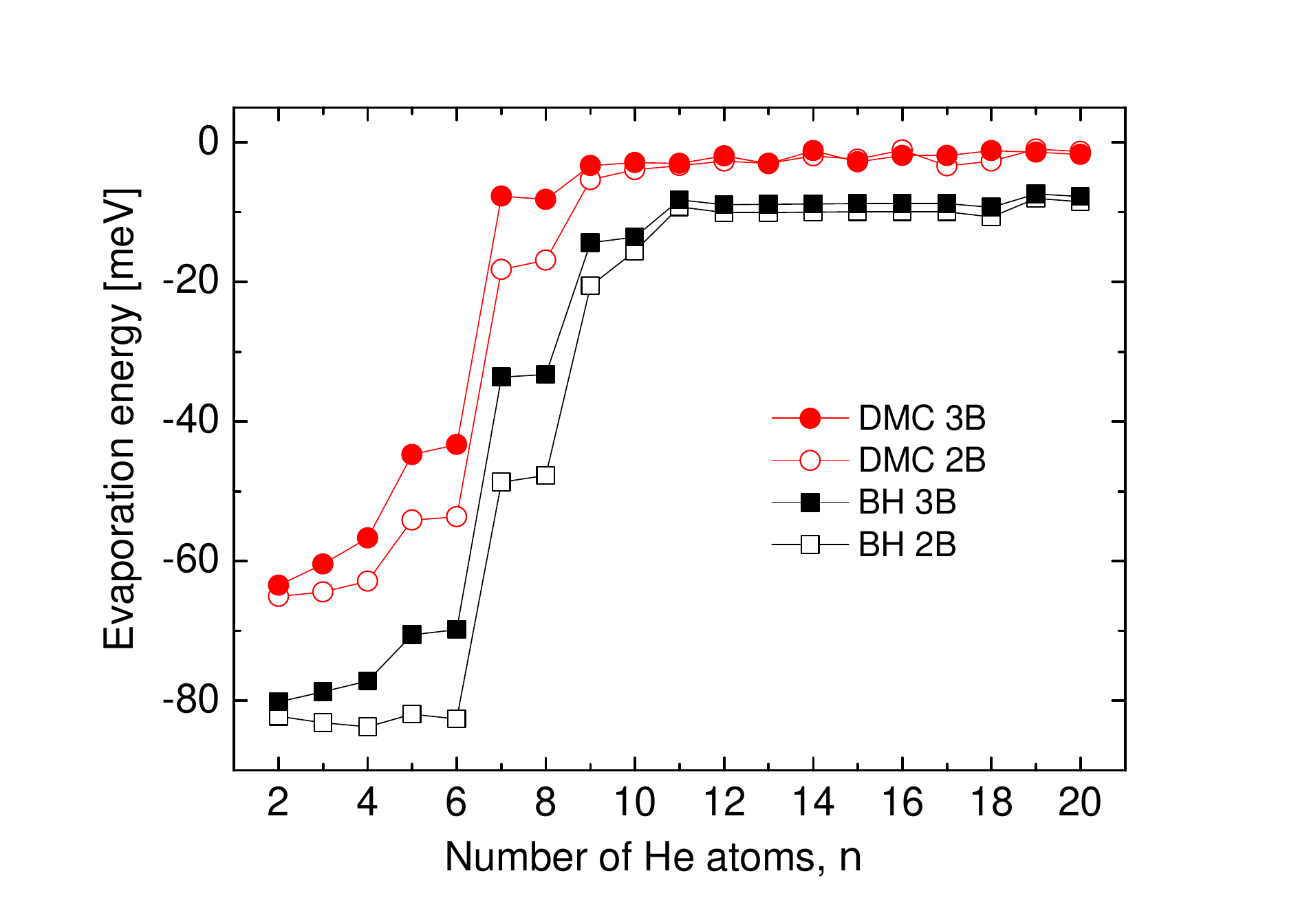} 
   \caption{Evaporation energy as a function of the number of He atoms $n$ for He$_n$Li$^+$ clusters calculated by means of the BH (black squares) and DMC (red circles) approaches using merely a pairwise description (empty symbols) and including 3B effects (solid symbols). See text for further details.} 
   \label{fig:2Bvs3B}
\end{figure}

In an attempt to test the effect of 3B terms on our present results we introduce a conveniently damped induced dipole-induced dipole interaction contributions as in Ref.\cite{LWHS:JCP16} and calculate the corresponding evaporation energies. 
The comparison of this magnitude as a function of the number of He atoms obtained by means of the present BH and DMC methods is shown in Figure \ref{fig:2Bvs3B}.
Some differences are certainly observed between those energies calculated only with the 2B pairwise description and those with the 3B terms included, especially for the smallest clusters $n < 8$. 
Beyond that size evaporation energies are practically the same regardlessly the potential interaction employed.
However the qualitative trend is the same for both the classical and the QM results in the figure. 
In addition, the second difference energies (not shown here) calculated with the 3B effects do not exhibit substantially different features in comparison with Figure \ref{fig:2ndE}, and in particular, no new peaks are seen. 
This suggests that contributions from terms beyond a mere 2B description, being significant in terms of the absolute energies of the clusters, do not improve in essence the comparison between theoretical and experimental results shown in this work.

\section{Conclusions}
\label{sec_conclu}

We have studied the formation and structures of Atkins snowballs around Li$^+$ ions using high resolution mass spectrometry and a number of different theoretical methods. The experiments show a series of particularly abundant He$_n$Li$^+$ complexes at $n=2$, 6, 8, and 14, as well as some weaker features such as minima at $n=21$, 24, 27, and 28.

The theoretical results show that classical approaches such as Basin-Hopping (BH) predicts qualitatively similar cluster properties as the quantum mechanical approaches of Quantum Free Energy (QFE), Diffusion Monte Carlo (DMC), and Path Integral Monte Carlo (PIMC). The simulations identify three particularly stable He shells surrounding the Li$^+$ ions for $n=4$, 6, 8, and a slightly weaker structure at 10. The sizes of $n=6$ and $n=8$ line up well with the experimental findings, suggesting that these features in the experimental mass spectrum are the results of these clusters stable octahedral and parallel square geometries, respectively. A larger magic structure observed for He$_{14}$Li$^+$ is observed in the experiments, but not in the simulations, could indicate the formation of multiple rigid and nested solvation shells.

\section*{Acknowledgments}

This work has been funded by the MINECO with Grants FIS2014-51993-P, FIS2016-79596-P, FIS2017-84391-C2-2-P and FIS2017-83157-P, the Austrian Science Fund FWF (projects P26635 and W1259), and the Swedish Research Council (Contract No.\ 2016-06625). JOZ would like to thank Programa Operativo de Empleo  Juvenil  2014--2020 of Fondo Social Europeo.



\balance


\bibliography{biblio_theo.bib} 
\bibliographystyle{rsc} 

\end{document}